\pdfoutput=1
\documentclass[reprint,aps,prl,twocolumn,superscriptaddress]{revtex4-1}
\usepackage{graphicx}
\usepackage{dcolumn}
\usepackage{bm}
\usepackage{amsmath}
\usepackage[english]{babel}
\usepackage[utf8]{inputenc}
\usepackage{color}
\begin{document}

\title{Low percolation density and charge noise with holes in germanium}

\author{M. Lodari}
\thanks{These authors contributed equally to this work}
\affiliation{QuTech and Kavli Institute of Nanoscience, Delft University of Technology, PO Box 5046, 2600 GA Delft, The Netherlands}
\author{N. W. Hendrickx}
\thanks{These authors contributed equally to this work}
\affiliation{QuTech and Kavli Institute of Nanoscience, Delft University of
Technology, PO Box 5046, 2600 GA Delft, The Netherlands}
\author{W. I. L. Lawrie}
\affiliation{QuTech and Kavli Institute of Nanoscience, Delft University of
Technology, PO Box 5046, 2600 GA Delft, The Netherlands}
\author{T. -K. Hsiao}
\affiliation{QuTech and Kavli Institute of Nanoscience, Delft University of
Technology, PO Box 5046, 2600 GA Delft, The Netherlands}
\author{L. M. K. Vandersypen}
\affiliation{QuTech and Kavli Institute of Nanoscience, Delft University of
Technology, PO Box 5046, 2600 GA Delft, The Netherlands}
\author{A. Sammak}
\affiliation{QuTech and Netherlands Organisation for Applied Scientific Research (TNO), Stieltjesweg 1, 2628 CK Delft, The Netherlands}
\author{M. Veldhorst}
\affiliation{QuTech and Kavli Institute of Nanoscience, Delft University of Technology, PO Box 5046, 2600 GA Delft, The Netherlands}
\author{G. Scappucci}
\email{g.scappucci@tudelft.nl}
\affiliation{QuTech and Kavli Institute of Nanoscience, Delft University of Technology, PO Box 5046, 2600 GA Delft, The Netherlands}

\date{\today}

\begin{abstract}
We engineer planar Ge/SiGe heterostructures for low disorder and quiet hole quantum dot operation by positioning the strained Ge channel 55~nm below the semiconductor/dielectric interface. In heterostructure field effect transistors, we measure a percolation density for two-dimensional hole transport of $2.1\times10^{10}~\text{cm}^{-2}$, indicative of a very low disorder potential landscape experienced by holes in the buried Ge channel. These Ge heterostructures support quiet operation of hole quantum dots and we measure charge noise levels that are below the detection limit $\sqrt{S_\text{E}}=0.2~\mu \text{eV}/\sqrt{\text{Hz}}$ at 1 Hz. These results establish planar Ge as a promising platform for scaled two-dimensional spin qubit arrays.  

\end{abstract}

\maketitle

The promise of quantum information with quantum dots\cite{DiVincenzo2000} has led to extensive studies for suitable quantum materials. While initial research mainly focused on gallium arsenide heterostructures because of its extremely low level of disorder, hyperfine interaction with nuclear spins has limited the quantum coherence of spin qubits\cite{Petta2005, Koppens2006}. Instead, silicon naturally contains only few non-zero nuclear spin isotopes and can furthermore be isotopically enriched, such that quantum coherence can be maintained for very long times\cite{veldhorst2014, Muhonen2014}. However, the relatively large effective mass and the presence of valley states complicates scalability\cite{zwanenburg_silicon_2013} and warrants the search for alternative quantum materials.  

Germanium has prospects to overcome these challenges and is rapidly emerging as a unique material for quantum information\cite{scappucci_germanium_2020}. Holes in germanium exhibit strong and tunable spin-orbit coupling allowing for fast and all-electrical control of spin qubits\cite{Bulaev2005, Bulaev2007, Watzinger2018, hendrickx_fast_2020}. The light effective mass along the Ge quantum well (QW) interface induces large orbital energy spacing for quantum dots and thereby relaxes the lithographic fabrication requirements\cite{lodari_light_2019}. In addition, ohmic contacts can be made to metals, which enabled the development of hybrid superconductor-semiconductor circuits\cite{hendrickx_gate-controlled_2018, hendrickx_ballistic_2019, vigneau_germanium_2019}, and promises novel approaches for long-range qubit links to engineer scalable qubit tiles\cite{Vandersypen2017InterfacingCoherent}.  

Importantly, Ge QWs can be engineered in silicon-germanium (Ge/SiGe) heterostructures\cite{sammak_shallow_2019} that are fabricated using techniques compatible with existing semiconductor manufacturing\cite{Pillarisetty2011Academic}, which facilitates large scale device integration. These advances enabled to define stable quantum dots\cite{hendrickx_gate-controlled_2018}, to operate quantum dot arrays\cite{lawrie_quantum_2020}, to realize single hole spin qubits\cite{hendrickx_single-hole_2020} with long spin life-times\cite{lawrie_spin_2020}, and to demonstrate full two-qubit logic\cite{hendrickx_fast_2020}. In all these experiments, the Ge QW was located remakarbly close to semiconductor/dielectric interface at a depth of only 22 nm\cite{sammak_shallow_2019}. While this shallow heterostructure showed an ultra-high maximum mobility exceeding $5\times10^{5}~\text{cm}^2/\text{Vs}$, possibly due to passivation of surface impurities by tunneled carriers from the QW, a rather high percolation density $p_p = 1.2\times10^{11}~\text{cm}^{-2}$ was measured. This value is similar to the values reported for Si metal-oxide semiconductor field effect transistors\cite{tracy2009observation,kim_annealing_2017,sabbagh_quantum_2019} and about twice the value reported in Si/SiGe QWs\cite{mi_magnetotransport_2015,wuetz_multiplexed_2020}. Since the percolation density characterizes disorder at low densities, which is the typical regime for quantum dot operation, a significant development is still needed to make undoped Ge/SiGe heterostructures compatible with existing architectures for large-scale quantum information processing with quantum dots, all relying on highly uniform qubits that exhibit extremely low noise\cite{Vandersypen2017InterfacingCoherent,li_crossbar_2018}. 

Here, we demonstrate planar Ge/SiGe heterostructures with very low levels of disorder and charge noise, setting new benchmarks for semiconductor materials for spin qubits. We quantify disorder beyond the metric of maximum mobility and focus on the percolation density, the single-particle relaxation time ($\tau{_q}$), which measures the time for which a momentum eigenstate can be defined even in the presence of scattering\cite{das_sarma_single-particle_1985}, and we report the associated quantum mobility $\mu{_q}=e\tau{_q}/m^*$\cite{harrang_quantum_1985}, with $e$ the elementary charge and $m^*$ the effective mass. By increasing the separation between the QW and the semiconductor/oxide interface to 55~nm, both $p_p$ and $\mu{_q}$ improve, and we find percolation densities as low as $p_p=2.1\times10^{10}~\text{cm}^{-2}$ and quantum mobilities as high as $\mu{_q}=2.5\times10^{4}~\text{cm}^{2}/\text{Vs}$. We introduce a method to measure charge noise in gate-defined quantum dots by scanning over Coulomb peaks to discriminate between measurement and device noise. We find that charge noise can be below our detection limit of $\sqrt{S_\text{E}}=0.2~\mu \text{eV}/\sqrt{\text{Hz}}$ at 1 Hz, about an order of magnitude less than previously reported for germanium quantum dots\cite{hendrickx_gate-controlled_2018}.  

\begin{figure}[!ht]
	\includegraphics[width=85mm]{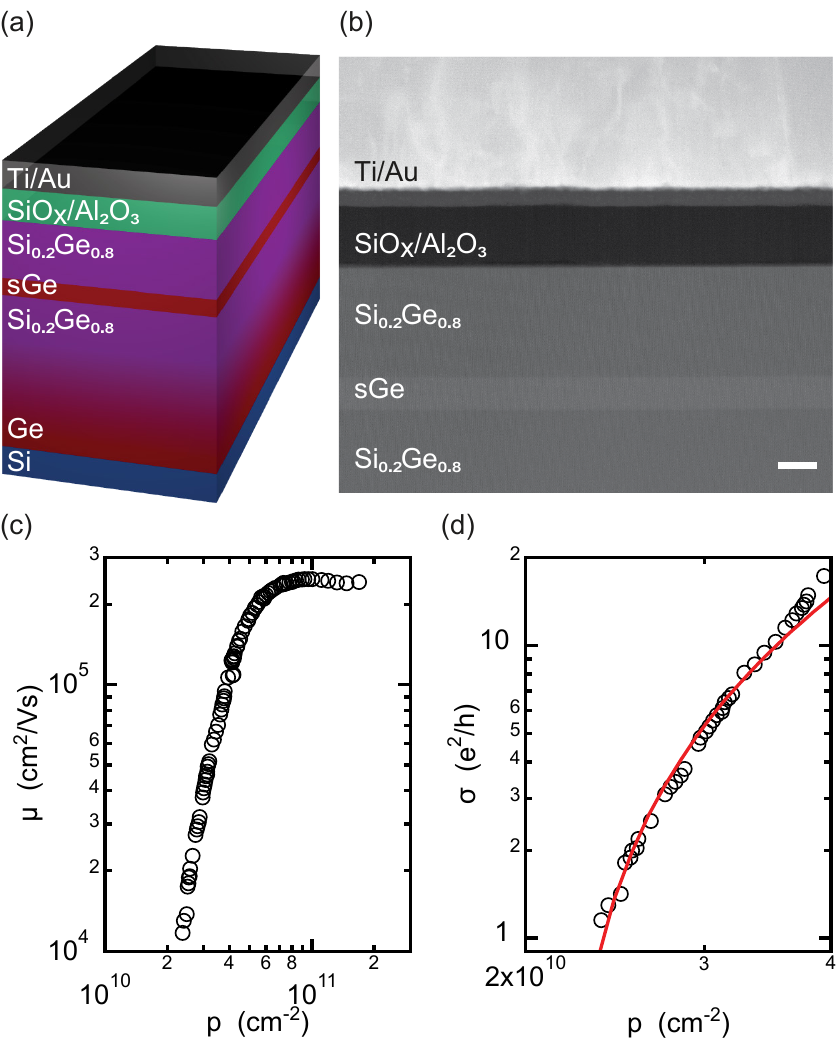}%
	\caption{(a) Schematic of a Ge/SiGe heterostructure field effect transistor. The strained Ge (sGe) quantum well is grown coherent to a strain-relaxed Si$_{0.2}$Ge$_{0.8}$ layer obtained by reverse grading. A Si$_{0.2}$Ge$_{0.8}$ barrier separates the QW from the dielectric stack---native silicon oxide followed by Al$_2$O$_3$--- and from the Ti/Au metallic gate metal. (b) High angle annular dark field scanning transmission electron image of the active layers of the Ge/SiGe heterostructure field effect transistor. Scale bar is 20~nm. (c) Mobility $\mu$ and (d) conductivity $\sigma_{xx}$ as a function of density $p$ at a temperature $T$ = 1.7~K. The red line in (d) is a fit to percolation theory in the low density regime.}
\label{fig:MAT}
\end{figure}

We grow Ge/SiGe heterostructures by reduced-pressure chemical vapor deposition on a Si(001) wafer and fabricate Hall-bar shaped heterostructure field effect transistors (H-FETs) for magnetotransport characterization by four-probes low-frequency lock-in techniques as described in Ref~\cite{sammak_shallow_2019}. Figure~\ref{fig:MAT}(a) shows a cross‐section schematic of the H-FET in the channel region. Figure~\ref{fig:MAT}(b) shows a high angle annular dark field scanning transmission electron (HAADF-STEM) image of the active layers of the H-FET, with no visible defects or dislocations. The strained Ge QW is uniform, has a constant thickness of 16~nm, and is separated from the SiO$_x$/Al$_2$O$_3$ dielectric stack by a Si$_{0.2}$Ge$_{0.8}$ barrier. We chose a Si$_{0.2}$Ge$_{0.8}$ barrier thickness $t=55~\text{nm}$ to suppress surface tunneling from the strained Ge QW\cite{su_effects_2017}, whilst achieving a sharp confinement potential for quantum dots. We achieve smooth interfaces between the Ge QW and nearby Si$_{0.2}$Ge$_{0.8}$ and between then Si$_{0.2}$Ge$_{0.8}$ barrier and the dielectric, highlighting the high-quality of epitaxy and device processing.

Applying a negative bias to the Ti/Au gate induces a two-dimensional hole gas and controls the carrier density in the QW. Figure~\ref{fig:MAT}(c) shows the transport mobility $\mu$ as a function of density $p$. The mobility increases steeply to 1$\times10^{5}~\text{cm}^{2}/\text{Vs}$ in the low-density range (2.4 to 3.9$\times10^{10}~\text{cm}^{-2}$) due to increased screening of scattering from remote charged impurities, likely at the semiconductor/dielectric interface. At higher density, the mobility also becomes limited by short-range scattering from impurities within or near the QW and saturates, reaching a maximum value of $2.5\times10^{5}~\text{cm}^{2}/\text{Vs}$ at a density of $9.2\times10^{10}~\text{cm}^{-2}$. The saturation of mobility upon increasing density indicates that surface tunneling is suppressed in this H-FET. In shallow Ge/SiGe heterostructures, an upturn in $\mu$ vs. $p$ dependence was observed above $p=3\times10^{11}~\text{cm}^{-2}$ instead, with no sign of saturation\cite{sammak_shallow_2019}. Figure~\ref{fig:MAT}(d) shows the conductivity $\sigma$ as a function of density $p$. We extract a percolation density of $p_p=2.14\times10^{10}~\text{cm}^{-2}$ by fitting $\sigma$ in the low density regime to percolation theory\cite{tracy2009observation,kim_annealing_2017,sammak_shallow_2019}. For measurements across two H-FETs fabricated on the same wafer we obtain a weighted average percolation density $<p_p>=(2.17\pm0.02)\times$10$^{10}~\text{cm}^{-2}$, pointing to uniform heterostructure deposition across the wafer and fabrication process. The obtained $p_p$ is indicative of very low disorder at low density, which is the typical condition for quantum dot operation, representing a $\approx5\times$ improvement compared to previous heterostructures supporting Ge spin qubits\cite{sammak_shallow_2019}, and setting a new benchmark for group-IV materials that have practical use for spin qubits. 

We further characterize disorder in the Ge H-FET by measuring the single-particle relaxation time $\tau_q$ and the associated quantum mobility $\mu_q$. $\tau_q$ determines the quantum level broadening $\Gamma = \hbar/2\tau_q$ of
the momentum eigenstates and is affected by all scattering events. This is distinct from the transport scattering time $\tau_t$, which instead is unaffected by forward scattering\cite{das_sarma_single-particle_1985} and determines the conductivity and the classical mobility $\mu=e\tau_t/m^*$. As such we argue that $\mu_q$ is a disorder qualifier less forgiving than $\mu$ and in principle is more informative of the qubit surrounding environment, since $\mu_q$ does not exclude a priori any source of scattering, which in turn might degrade qubit performance.

\begin{figure}[!ht]
	\includegraphics[width=85mm]{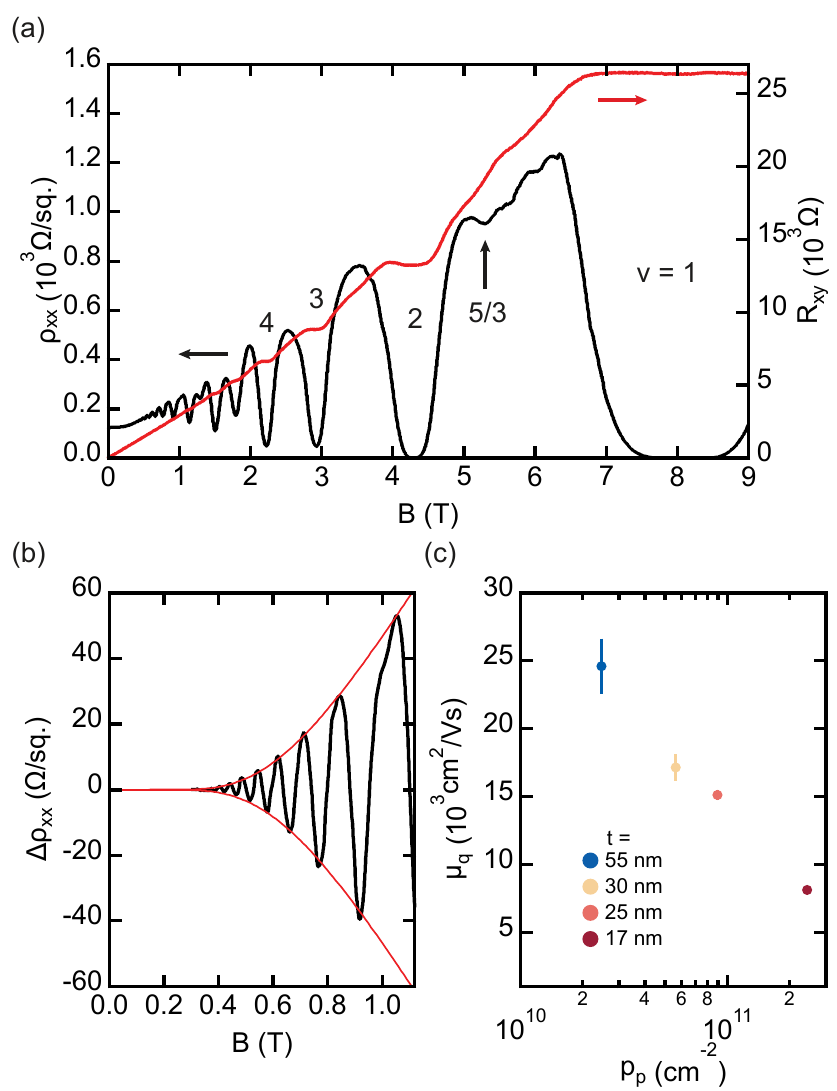}%
	\caption{(a) Longitudinal resistivity ($\rho_{xx}$, black curve) and transverse Hall resistance (R$_{xy}$, red curve) as a function of perpendicular magnetic field $B$ measured at a density $p$ = $2.1\times10^{11}$ cm$^{-2}$. The first four Landau level integer filling factors ($\nu$) are labeled, along with the 5/3 fractional state . (b) Low-magnetic field oscillation amplitude ($\Delta \rho_{xx}$, black curve) as a function of $B$ after polynomial background subtraction and theoretical fit of the envelope (red curve) to evaluate the single-particle relaxation time $\tau_q$. (c) Quantum mobility ($\mu_q$) as a function of percolation density measured in heterostructures with barrier thickness $t$ in the range of 17-55 nm.}
\label{fig:QUANTUM}
\end{figure}

To measure $\tau_q$ and $\mu_q$ we probe the disorder-induced broadening of the 2DHG Landau levels in magnetotransport. Figure~\ref{fig:QUANTUM}(a) shows the longitudinal resistivity $\rho_{xx}$ and transverse Hall resistance $R_{xy}$ as a function of $B$ at a fixed density corresponding to the maximum transport mobility. We observe Shubnikov-de Haas oscillations above $B= 0.37~\text{T}$ and Zeeman splitting above $B= 0.83~\text{T}$, from which we estimate an effective $g^* = 12.7$ following the methodology in Ref.~\cite{sammak_shallow_2019}. The oscillation minima go to zero above $B= 4.3~\text{T}$, signaling high quality magnetotransport from a single high-mobility subband corresponding to the heavy hole fundamental state in the Ge QW. $R_{xy}$ develops flat plateaus corresponding to oscillation minima in $\rho_{xx}$, due to the integer quantum Hall effect. Signatures of the $\nu = 5/3$ fractional state are visible both in $\rho_{xx}$ and $R_{xy}$, indicating a robust energy gap that survives thermal broadening at 1.7~K. 

Figure~\ref{fig:QUANTUM}(b) shows the low-field oscillation amplitude $\Delta\rho_{xx} = (\rho_{xx}-\rho_0)$ as a function of perpendicular magnetic field $B$, where $\rho_0$ is the $\rho_{xx}$ value at $B = 0$. We estimate a single-particle relaxation time $\tau_q = 0.87~\text{ps}$ from a fit of the Shubnikov-de Haas oscillation envelope to the function $\Delta\rho_{xx} \approx \rho_0B^{1/2} \chi/\sinh(\chi) \exp(-\pi / \omega_c \tau_q)$, where $\chi = 2\pi^2 k_B T/\hbar\omega_c$, $k_B$ is the Boltzmann constant, $\hbar$ is the Planck constant and $\omega_c$ is the cyclotron frequency (Fig.~\ref{fig:QUANTUM}(b), red curve)\cite{bauer_low-temperature_1972}\footnote{For the analysis of $\tau_q$ and $\mu_q$ in Fig.~\ref{fig:QUANTUM}(b) and (c), we extrapolate the effective mass $m^*$ from Ref.~\cite{lodari_light_2019} at the relevant density. Specifically, for the 55-nm-deep quantum well discussed here, we assume $m^* = 0.062 \times m_0$ at the saturation density $p_{sat}=2.1\times10^{11}~\text{cm}^{-2}$}. Correspondingly, we estimate $\Gamma = 377~\mu\text{eV}$. This is $\approx 4\times$ smaller than $\Gamma$ at a comparable $p$ in a shallow QW positioned 17~nm below the surface, signaling that disorder is greatly reduced in the heterostructure detailed in this work. We find a Dingle ratio $\tau_t/\tau_q=10$, which is $\approx 3\times$ smaller compared to shallower QWs\cite{sammak_shallow_2019}, confirming that long-range scattering is reduced, as expected from the $\mu$ dependence on $p$ in Fig.~\ref{fig:MAT}(c).

In Figure~\ref{fig:QUANTUM}(c) we show the quantum mobility $\mu_q$ as a function of the percolation density $p_p$ measured for QWs positioned at different distance $t$ from the semiconductor/dielectric interface. For each heterostructure, $\mu_q$ is evaluated at saturation density $p_{sat}\sim1/t$\cite{lodari_light_2019}. We observe a clear correlation: as the QW is separated from the impurities at the semiconductor/dielectric interface, both our disorder qualifiers $p_p$ and $\mu_q$ improve and reach best values in the heterostructure with $t=55~\text{nm}$. The observed correlation also implies that  percolation density, which may be measured at higher temperatures and more easily than Shubnikov-de Haas oscillations, is sufficient to provide a fast feedback loop on heterostructure growth and device processing.

\begin{figure}[!ht]
	\includegraphics[width=85mm]{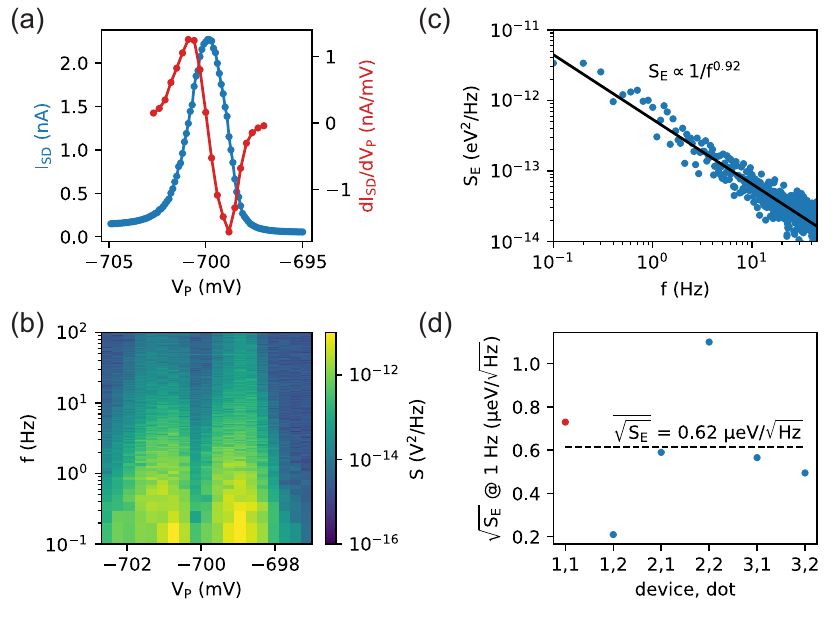}%
	\caption{(a) Source-drain current $I_{SD}$ (blue) through quantum dot 1 of device 1 as well as the numerical derivative (red) related to the sensitivity of the source-drain current to electric field variations.
	        (b) Frequency dependence of the power spectrum density of $I_{SD}$, for different plunger gate voltages $V_{P}$. Each trace consists of 10 averaged 10-second samples of the source-drain current.
	        (c) Power spectrum density of the noise picked up by quantum dot 1 of device 1, for $V_P = -698.8~\text{mV}$. Solid line corresponds to apparent linear fit to the data, yielding a slope of -0.92.
	        (d) The charge noise measured at $f = 1~\text{Hz}$ for six different quantum dots in three different devices. The point in red corresponds to the data in panels (a-c). Dashed line indicates the mean value across all quantum dots.
	        }
\label{fig:chargenoise}
\end{figure}

We now move on to the formation of quantum dots in this platform. We fabricate six quantum dots in three different devices, all consisting of a set aluminium Ohmic leads, as well as two metallic gate layers defining the quantum dots\cite{lawrie_quantum_2020}. We operate the quantum dots in transport mode by applying a bias voltage across the quantum dot Ohmic leads and measuring the resulting current for each dot. In Fig.~\ref{fig:chargenoise}(a) we measure the source-drain current $I_\text{SD}$ in blue as a function of the applied plunger gate voltage $V_\text{P}$ and a typical Coulomb peak in the device conductance can be observed. 

To qualify the quantum dot environment, we measure the charge noise picked up by a single quantum dot. A 100-s long trace of $I_\text{SD}$ is acquired and split into 10 segments of equal lengths. The power spectrum density of the noise $S$ is obtained by averaging the discrete Fourier transform of the 10 segments. In order to distinguish noise caused by the measurement equipment from charge noise acting on the quantum dot, we repeat the same measurement for different quantum dot plunger gate voltages spanning a full Coulomb peak. Figure~\ref{fig:chargenoise}(a) shows $I_\text{SD}$ (blue), as well as the numerical derivative $\delta I_\text{SD}/\delta{V_P}$ (red) indicating the sensitivity of the source-drain current to electric field variations, for all gate voltages where charge noise measurements are performed. In Figure~\ref{fig:chargenoise}(b) we show the noise spectrum density as a function of both $V_P$ as well as frequency $f$. A clear increase of $S$ can be observed on the flanks of the Coulomb peak, corresponding to the points of highest sensitivity. At the top of the Coulomb peak, where the local derivative of the source drain current is close to zero, the noise spectral density drops. This is a clear indication that the measured spectrum originates in the environment of the quantum dot and not the measurement equipment or other noise sources such as tunnelling noise\cite{jung_background_2004, freeman_comparison_2016}. We argue that solely comparing the noise spectrum at the flank of a Coulomb peak to the noise spectrum in Coulomb blockade is not sufficient, as the noise floor of a transimpedance amplifier often strongly depends on the impedance of the load. By moving from Coulomb blockade to the flank of a Coulomb peak, the device impedance can typically change from $R_\text{block} > 100~\text{G}\Omega$ to $R_\text{transport} < 1~\text{M}\Omega$, thereby rendering a comparison of the two noise spectra invalid. The difference in device impedance between the flank and top of a Coulomb peak is typically less than an order of magnitude and is therefore a good indicator of the source of the observed noise spectrum. Figure \ref{fig:chargenoise}(c) shows the equivalent detuning noise spectral density $S_\text{E}$ measured at $V_\text{P} = -698.8$ mV, using a lever arm of $\alpha \approx 0.1$ as obtained from Coulomb diamond measurements for each dot. The spectrum follows an approximate $1/f$ trend at low frequencies\cite{freeman_comparison_2016}, allowing us to extract an equivalent detuning noise at 1 Hz. We perform charge noise measurements on all six quantum dots and the results are presented in Fig.~\ref{fig:chargenoise}(d). The average detuning noise at 1 Hz is $\overline{\sqrt{S_\text{E}}}=0.6~\mu\text{eV}/\sqrt{\text{Hz}}$, with the lowest value being limited by our measurement setup at $0.2~\mu\text{eV}/\sqrt{\text{Hz}}$. This is several times smaller than the charge noise $\sqrt{S_\text{E}}=1.4~\mu \text{eV}/\sqrt{\text{Hz}}$ reported in shallower 22-nm-deep Ge QWs\cite{hendrickx_gate-controlled_2018}. Moreover, the lowest charge noise values reported here compare favourably to other material systems, $0.5~\mu \text{eV}/\sqrt{\text{Hz}}$ for Si/SiO$_2$\cite{connors_low-frequency_2019}, $0.8~\mu \text{eV}/\sqrt{\text{Hz}}$ for Si/SiGe\cite{freeman_comparison_2016}, $\sim1~\mu \text{eV}/\sqrt{\text{Hz}}$ for InSb\cite{jekat_exfoliated_2020} or $7.5~\mu \text{eV}/\sqrt{\text{Hz}}$ for GaAs\cite{basset_evaluating_2014}, thereby setting the benchmark for semiconductor quantum dots.

In summary, we have engineered planar Ge/SiGe heterostructures for low disorder and quiet quantum dot operation. We measure  a percolation density for two-dimensional hole conduction $p_p=2.14\times 10^{10}~\text{cm}^{-2}$. At such low carrier density, measurements might be limited by the contact resistance leaving room for further improvement. In quantum dots, charge noise is below the detection limit $\sqrt{S_\text{E}}=0.2~\mu \text{eV}/\sqrt{\text{Hz}}$ at 1 Hz of our setup. These results mark a significant step forward in the germanium quantum information route.

\section{Acknowledgments}
We acknowledge support through a
FOM Projectruimte of the Foundation for Fundamental Research on Matter (FOM), associated with the Netherlands Organisation for Scientific Research (NWO). Research was sponsored by the Army Research Office (ARO) and was accomplished under Grant No. W911NF- 17-1-0274. The views and conclusions contained in this document are those of the authors and should not be interpreted as representing the official policies, either expressed or implied, of the Army Research Office (ARO), or the U.S. Government. The U.S. Government is authorized to reproduce and distribute reprints for Government purposes notwithstanding any copyright notation herein. We acknowledge the Quantera ERA-NET Cofund in Quantum Technologies implemented within the European Union’s Horizon 2020 Programme

\vspace{\baselineskip}
Data sets supporting the findings of this study are available at 10.4121/uuid:70cf99ac-5914-4381-abfa-8f9eed7004fd.

\bibliography{main.bib}

\end{document}